\documentstyle[12pt]{article}

\newcommand{\slp}{\raise.15ex\hbox{$/$}\kern-.57em\hbox{$\partial$}}
\newcommand{\sla}{\raise.15ex\hbox{$/$}\kern-.57em\hbox{$a$}}
\newcommand{\slA}{\raise.15ex\hbox{$/$}\kern-.57em\hbox{$A$}}
\newcommand{\slB}{\raise.15ex\hbox{$/$}\kern-.57em\hbox{$B$}}
\newcommand{\slb}{\raise.15ex\hbox{$/$}\kern-.57em\hbox{$b$}}
\newcommand{\slW}{\raise.15ex\hbox{$/$}\kern-.57em\hbox{$W$}}
\newcommand{\dA}{\sqcup\!\!\!\!\!\sqcap}
\newcommand{\da}{\sqcup\!\!\!\!\sqcap}
\newcommand{\be}{\begin{equation}}
\newcommand{\ee}{\end{equation}}
\newcommand{\bear}{\begin{eqnarray}}
\newcommand{\ear}{\end{eqnarray}}

\newcommand{\tr}{\rm tr}
\begin{document}
%\begin{titlepage}
\begin{flushright}
HD--THEP--95--34\\
\end{flushright}
%\quad\\
%\renewcommand{\baselinestretch}{1.6}
\vskip1.5cm
\begin{center}
{\bf\LARGE BRST Cohomology and Vacuum Structure}\\
\vskip.3cm
{\bf\LARGE of Two-Dimensional Chromodynamics}\\
\vspace{1cm}
E. Abdalla\footnote{Work supported by Alexander von
Humboldt Stiftung. On leave from the University of Sao Paulo,
Brazil.}
 and K. D. Rothe\\
\bigskip
Institut  f\"ur Theoretische Physik\\
Universit\"at Heidelberg\\
Philosophenweg 16, D-69120 Heidelberg\\
\end{center}
\vspace{2.0cm}
\begin{abstract}\noindent
Using a formulation of $QCD_2$ as a perturbed conformally
invariant theory involving fermions, ghosts, as well as positive and
negative level Wess-Zumino-Witten fields, we show that the BRST
conditions become restrictions on the conformally invariant
sector, as described by a G/G topological theory.  By solving
the corresponding cohomology problem we are led to a finite set
of vacua. For $G=SU(2)$ these vacua are two-fold degenerate.
 \end{abstract}
%\end{titlepage}
\newpage
\pagestyle{plain}
\setcounter{equation}{0}

Quantum Chromodynamics in 1+1 dimensions ($QCD_2$) has been subject
of numerous investigations in the past twenty five years \cite{AAR},
\cite{AA2}. However, unlike its abelian counterpart, the exactly
soluble Schwinger model \cite{Schw}, one had up to recently
no hint at its exact integrability. Moreover, traditional topological
arguments based on instantons suggest that the vacuum of $QCD_2$ is
unique, unlike the case of the Schwinger model, where this vacuum is
known to be infinitely degenerate. However, arguments have been presented
\cite{Wi} in favor of the existence of a discrete, but finite set
of $QCD_2$ vacua in higher representations of the fermions.

The recent formulation \cite{AA1} of $QCD_2$ as a perturbed
conformally invariant Wess-Zumino-Witten-type theory \cite{WZW}
turns out to provide an appropriate starting point for a dynamical
investigation of the physical Hilbert-space structure of $QCD_2$.
The fundamental framework is provided by the BRST analysis of $QCD_2$
in this formulation, as carried out in ref. \cite{CRS}. In particular it
will be our aim to investigate the possible existence of
degeneracy of the $QCD_2$ vacuum.In this respect it will be useful to
point out the parallelisms with the Schwinger model in the decoupled
formulation.We shall show that the conformally invariant
sector of $QCD_2$ is
described by a level-one G/G topological field theory, thus allowing
for a complete classification of the
ground states. By explicitly solving
the cohomology problem for G=SU(2), we find that the vacuum
is two-fold degenerate in the left and right moving sector, respectively.

%\section{BRST Symmetry and Constraint Structure}
%\setcounter{equation}{0}
In order to provide the necessary framework, we briefly
review the essential results of refs. \cite{AA1},\cite{CRS}.

In the light-cone gauge $A_+=0$, the $QCD_2$-partition function
reads\footnote{Our conventions are: $\chi={\chi_1 \choose \chi_2},
\ \gamma_+=
2{0\,1 \choose 0\,0}, \gamma_-=2{0\,0 \choose 1\,0}, A_\pm=A_0\pm A_1,
\ \partial_\pm=\partial_0\pm\partial_1,\ A_\mu=A^a_\mu t^a$, etc.,
with $[t^a,t^b]=ifabct^c$ and $tr(t^at^b)=\delta^{ab},
fabcfabd=\frac{1}{2} C_V\delta cd$.}
\be\label{2.1}
Z=\int{\cal D} A_-\int{\cal D}\chi_1^{(0)}{\cal D}\chi_1^{\dagger(0)}\int
{\cal D}\psi_2
{\cal D}\psi_2^\dagger\int{\cal D} b_-{\cal D} c_-e^{iS_{GF}}\ee
with the corresponding gauge-fixed Lagrangian \cite{CRS}
\be\label{2.2}
{\cal L}_{GF}=tr\frac{1}{8}(\partial_+A_-)^2+\chi_1^\dagger i
\partial_+\chi_1+\psi^\dagger_2 iD_-\psi_2+tr(b_-i\partial_+c_-)\ee
where $b_-,c_-$ are Grassman-valued ghost fields arising from the
gauge-fixing condition.

\bigskip
\noindent{\it Local decoupled formulation of $QCD_2$}

\medskip
Making the change of variable
\bear\label{2.3}
A_-&=&\frac{i}{e}V\partial_- V^{-1}\nonumber\\
\psi_2&=& V\chi_2\ear
one is led to the factorized partition function \cite{AA1,CRS}
\be\label{2.4}
Z=Z_F^{(0)}Z_{gh}^{(0)}Z_V\ee
where
\be\label{2.5}
Z_V=\int {\cal D} V e^{-i(c_V+1)\Gamma[V]-\frac{i}{8e^2}\int d^2 x\
tr[\partial_+(V\partial_-V^{-1})]^2}\ee
and
\bear\label{2.6}
Z_F^{(0)}&=&\int{\cal D}\bar\chi{\cal D}\chi e^{i\int d^2 x\bar\chi
i\partial
\!\!\!/\chi}\nonumber\\
Z^{(0)}_{gh}&=&\int {\cal D}b_\pm{\cal D}c_\pm e^{i\int d^2
x\tr[b_+i\partial_-c_++b_-i\partial_+c_-]}\ear
with $\Gamma[g]$ the WZW functional
\be\label{2.7}
\Gamma[g]=\frac{1}{8\pi}\int d^2x\ tr\partial^\mu g^{-1}\partial_\mu
g+\frac{1}{4\pi}\ tr\int^1_0 dr\int d^2 x\varepsilon^{\mu\nu}\tilde g^{-1}
\dot{\tilde g} \tilde g^{-1}\partial_\mu\tilde g\tilde
g^{-1}\partial_\nu\tilde g\ee
where $\tilde g(1,x)=g(x),\tilde g(0,x)=1\!{\rm l}$. The partition
function exhibits a BRST symmetry in the left- and right-moving sector,
implying the existence of conserved right- and left-moving BRST currents
 \cite{CRS}
\bear\label{2.8}
J^{(B)}_{\mp}&=&{\rm
tr}\left[c_\mp\Omega_\mp-\frac{1}{2}b_\mp\{c_\mp,c_\mp\}\right]\nonumber\\
&&\partial_\pm J_\mp^{(B)}=0\ear
where
\bear\label{2.9}
\Omega_-\equiv-\frac{1}{4e^2}{\cal D}_-(V)\partial_+(Vi\partial_-V^{-1})
&-&\left(\frac{1+C_V}{4\pi}\right)
Vi\partial_-V^{-1}\nonumber\\
&+&\chi_1\chi_1^\dagger+b_-\{c_-,c_-\}\approx 0\ear
and
\bear\label{2.10}
\Omega_+&\equiv&\frac{1}{4e}V^{-1}[\partial^2_+
(Vi\partial_-V^{-1})]V-\frac{1+C_V}
{4\pi}V^{-1}i\partial_+V\nonumber\\
&&+\chi_2\chi_2^\dagger+b_+\{c_+,c_+\}\approx 0\ear
are first-class constraints, with $\Omega_-\approx 0$
playing the role of the Gauss law. It is interesting to note that
\be\label{2.11}
\partial_-\Omega_+=-V^{-1}\partial_+\Omega_-V\ee
implying
\be\label{2.12}
{\cal D}_-\partial_+ A_- +(1+C_V)\frac{e^2}{\pi}A_-=0.\ee
Note that the term proportional to $C_V$ has been ignored in the
literature.

\bigskip
\noindent{\it Non-local decoupled formulation of $QCD_2$}

\medskip
The partition function (\ref{2.5}) involves 4th order derivatives.
 In order to reduce this order to 2nd order, we introduce an auxiliary
field $E$ via the identity
\be\label{2.13}
e^{\frac{i}{4e^2}\int\frac{1}{2}{\rm tr}[\partial_+(Vi\partial_-
V^{-1}]^2}=
\int{\cal D} E e^{-i\int\frac{1}{2}{\rm tr}\left[\frac{e^2}{\pi}
E^2+\frac{E}{\sqrt\pi}\partial_+(Vi\partial_-V^{-1})\right]}.\ee
Making the change of variable \cite{AA1},\cite{CRS}
\be\label{2.14}
E=\sqrt\pi\left(\frac{1+C_V}{2\pi}\right)\frac{1}{\partial_+}
(\beta^{-1} i\partial_+\beta)\ee
and making use of the Polyakov-Wiegmann identity \cite{PW}
\be\label{2.15}
\Gamma[gh]=\Gamma[g]+\Gamma[h]+\frac{1}{4\pi}\int d^2x\ {\rm tr}\left(
g^{-1}\partial_+ gh\partial_- h^{-1}\right)\ee
one arrives at the alternative representation \cite{AA1},\cite{CRS}
\be\label{2.16}
Z=Z_F^{(0)}Z_{gh}^{(0)}Z_{\tilde V}Z_\beta\ee
with
\be\label{2.17}
Z_{\tilde V}=\int{\cal D}\tilde V\exp\{-i(1+C_V)\Gamma[\tilde V]\}\ee
where $\tilde V=\beta V$, and
\be\label{2.18}
Z_\beta=\int{\cal
D}\beta\exp\left\{i\Gamma[\beta]+i\left(\frac{1+C_V}{2\pi}\right)^2
e^2\int\frac{1}{2}{\rm tr}\left[\partial_+^{-1}(\beta^{-1}\partial_+\beta)
\right]^2\right\},\ee
Note that the WZW action enters with negative level $-(1+C_V)$. This will
be very important in the following section.

There exist \cite{CRS} two BRST currents associated with
the partition function (\ref{2.16}):
\be\label{2.20}
\tilde J^{(B)}_\pm={\rm tr}\left[ c_\pm\tilde\Omega_\pm-\frac{1}{2}b_\pm
\{c_\pm,c_\pm\}\right]\ee
where
\bear
\tilde\Omega_-&\equiv&\chi_1\chi_1^\dagger+\{b_-^{(0)},c_-^{(0)}
\}-\frac{1+C_V}{4\pi}\tilde V i\partial_-\tilde V^{-1}
\approx 0\label{2.21}\\
\tilde\Omega_+&\equiv&\chi_2\chi_2^\dagger+\{b_+^{(0)},
c_+^{(0)}\}-\frac{1+C_V}{4\pi}\tilde V^{-1}i\partial_+\tilde V\approx
0\label{2.22}\ear
represent first class constraints. These shall play  a central role in the
characterization of the QCD$_2$ vacuum.

Finally let us rewrite $Z_\beta$ in (\ref{2.18}) in terms of an auxiliary
field $C_-$ as follows:
\be\label{2.23}
Z_\beta=\int{\cal D}\beta\int{\cal D}C_-e^{iS'}\ee
where
\be\label{2.24}
S'=\Gamma[\beta]+\int\left[\frac{1}{2}{\rm
tr}(\partial_+C_-)^2+\left(\frac{1+C_V}{2\pi}\right)e{\rm
tr}(C_-\beta^{-1} i\partial_+\beta)\right].\ee
By gauging the $gh-\tilde V-\beta$ sector following
method of ref. \cite{KS}, one discovers one further constraint
\be\label{2.25}
\left(\frac{1+C_V}{2\pi}\right)e\beta
C_-\beta^{-1}\left(\frac{1+C_V}{4\pi}\right)
\tilde V i\partial_-\tilde V^{-1}
-\frac{1}{4\pi}\beta i\partial_-\beta^{-1}+\{b_-,c_-\}\approx 0.\ee
As emphasized  in \cite{AA1}, this constraint is 2nd
class with respect to the constraints (\ref{2.21},
\ref{2.22}), and serves to determine the auxiliary field $C_-$.
\newpage

\bigskip
{\it The Schwinger model revisited}
\medskip

In order to gain some feeling for the constraints
(2.9), (2.10), (2.21) and
(2.22) it is useful to see what these constraints correspond to
in the $U(1)$ case.

In the $U(1)$ case, $C_V=0$ (corresponding to decoupled
Faddeev-Popov ghost from the outset). Parametrizing $V$ in
(\ref{2.3}) by
\be\label{3.1}
V=e^{i2\sqrt\pi\phi},\ee
the WZW functional $\Gamma[V]$ and Maxwell term in (2.5) reduce to
\be\label{3.3}
\Gamma[V]=\int\frac{1}{2}(\partial_\mu\phi)^2,\ \
S_{Max}=\left(\frac{2\sqrt\pi}{e}\right)^2
\int(\dA\phi)^2\ee
respectively, so that the partition function (\ref{2.4}) reads
\be\label{3.5}
Z=Z^{(0)}_F Z^{(0)}_{gh}
\int{\cal D}\phi e^{i\int d^2x\{-\frac{1}{2}\partial^\mu\phi\partial_\mu
\phi+\frac{\pi}{2e^2}(\da\phi)^2\}}\ee
Notice that $\phi$ is a negative metric field, corresponding
to the fact that the WZW action $\Gamma[V]$ enters in (\ref{2.5})
with negative level.

{}From (\ref{3.3}) follows the equation of motion
\be\label{3.6}
\dA(\dA+\frac{e^2}{\pi})\phi=0\ee
the constraints $\Omega_\pm$ (\ref{2.9}) and (\ref{2.10}) take
the form
\be\label{3.7}
\Omega_\pm=\frac{\sqrt\pi}{2e}\partial_\mu(\dA+\frac{e^2}{\pi})\phi
-e\bar\chi\gamma_\mu\chi\approx0\ee
In the spirit of \cite{KS} these constraints are obtained by the
gauging of the effective action in (3.5) as follows:
\bear\label{3.8}
&&\bar\chi i\slp\chi\to\bar\chi(i\slp+\slW)\chi\nonumber\\
&&-\int\frac{1}{2}(\partial_\mu\phi)^2\to-\frac{1}{2}\int(\partial
_\mu\phi+\frac{1}{2\sqrt\pi}W_\mu)^2\nonumber\\
&&\frac{\pi}{2e^2}\int(\dA\phi)^2\to\frac{\pi}{2e^2}\int\left[
\partial^\mu(\partial_\mu\phi+\frac{1}{2\sqrt\pi}W_\mu)\right]^2\ear
where $W_\mu$ is an external field. Parametrizing this field as
\be\label{3.9}
 W_\mu=\epsilon_{\mu\nu}\partial^\nu\psi+\partial_\mu
\zeta\ee
one finds
\bear\label{3.10}
&&-\frac{1}{2}\int(\partial_\mu\phi)^2\to-\frac{1}{2}\int(\partial_\mu
\tilde\phi)^2+\frac{1}{8\pi}\int(\partial_\mu\psi)^2\nonumber\\
&&\frac{\pi}{2e^2}\int(\dA\phi)^2\to\frac{\pi}{2e^2}\int(\dA
\tilde\phi)^2\ear
where $\tilde\phi=\phi+\frac{1}{2\sqrt\pi}\zeta$. The
$\psi$-dependent term cancels against the anomaly arising from
the fermionic integration, so that the gauged partition function
coincides with (\ref{3.5}). Following \cite{KS}, the variation
of the partition function with respect to $W_\mu$ then leads to the
constraints (\ref{3.7}). Notice that the Klein-Gordon operator
$(\dA+\frac{e^2}{\pi})$ projects out the massive mode of $\phi$
satisfying (\ref{3.6}), leaving one only with the massless mode.
Hence, (\ref{3.7}) corresponds to constraints on the massless
(conformally invariant)
sector of the theory. Indeed, both the curl and divergence of
$\Omega_\mu$ vanishes.

In the Schwinger model it is clear how to separate the massless
(negative metric) field from the massive (physical)
excitations. The procedure corresponds to the transition
from the local to the non-local formulation of section 2,
and consists in introducing the auxiliary field $E$ via (\ref{2.13})
with the parametrization (\ref{3.3}). This leads
to the new effective Lagrangian
\be\label{3.11}
\tilde{\cal L}=\bar\chi i\slp\chi
+b_+ i\partial_-c_++b_-i\partial_+ c_-
-\frac{1}{2}\partial_\mu \eta
\partial^\mu\eta+\frac{1}{2}\partial^\mu E\partial_\mu E
-\frac{e^2}{2\pi}E^2\ee
where
\be\label{3.12}
\eta=\phi-E\ee
The Lagrangian $\tilde{\cal L}$ plays the role of the ``non-local''
$QCD_2$ Lagrangian of section 2.

We see that $\eta$ is the negative metric, zero mass field in the usual
pa\-ra\-me\-tri\-zation of the Schwinger model.
In the non-abelian formulation,
the fields
$\beta, \tilde V$ and $V$ take the role of $E,\eta$ and $\phi$,
respectively, the correspondences being given by
$\beta=\exp(-2i\sqrt\pi E), \tilde V=\exp(2i\sqrt\pi\eta)$ as well
as (\ref{3.1}).

According to our discussion in section 2, we expect two
first-class constraints, and one second-class constraint.
The first two one obtains by gauging the fermion-eta sector
in a way analogous to the first two equations in (\ref{3.8}).
This leads to the first-class constraints
\be\label{3.13}
\tilde\Omega_\mu:=\bar\chi\gamma_\mu\chi-\frac{1}{2\sqrt\pi}
\partial_\mu\eta\approx0\ee
which replace the constraints (\ref{2.21}) and (\ref{2.22})
in the abelian case. Condition (\ref{3.13}), when implemented
on the physical states, is just the familiar requirement
\cite{LS}
\be\label{3.14}
\partial_\mu(\varphi+\eta)|phys\rangle=0\ee
where $\varphi$ is the ``potential'' of the free fermionic current
\be\label{3.15}
\bar\chi\gamma_\mu\chi=\frac{-1}{2\sqrt\pi}\partial_\mu\varphi.\ee
Condition (\ref{3.14}) characterizes the physical Hilbert space
of the Schwinger model, and in particular its ground state structure,
which turns out to be infinitely degenerate. In the non-abelian
case, this role is taken up by the constraints (\ref{2.21})
and (\ref{2.22}).

In the notation of this section, the partition function (\ref{2.23})
takes the form
\be\label{3.16}
Z_\beta=\int{\cal D}\beta{\cal D} C_-e^{i\int{\cal L}'}\ee
with
\be\label{3.17}
{\cal L}'=-\frac{1}{2}(\partial_\mu\eta)^2+\frac{1}{2}(\partial_\mu E)^2
+C_-\partial_+E+\frac{1}{2}(\partial_\mu C_-)^2\ee
The gauging of ${\cal L}'$, following the procedure of ref. \cite{KS}
corresponds to the replacement of ${\cal L}'$ by
\be\label{3.18}
{\cal L}_W={\cal L}'+W_+(\partial_-\eta+\partial_-E-\frac{e}{\sqrt\pi}
C_-)
\ee
with $W_+=\partial_+\phi$. Expression (\ref{3.18})
can be written as
\be\label{3.19}
{\cal L}''=-\frac{1}{2}(\partial_\mu\eta')^2+\frac{1}{2}(\partial_\mu
E')^2
+C_-\partial_+E'+\frac{1}{2}(\partial_\mu C_-)^2\ee
with $\eta'=\eta+\phi,\ E'=E+\phi$. Hence the partition
functions associated with ${\cal L}''$ and ${\cal L}'$ coincide, implying
the
constraint
\be\label{3.20}
-\frac{e}{\sqrt\pi}C_--\partial_-\eta+\partial_-E=0\ee
which is just the abelian version of (\ref{2.25}).

The physical Hilbert space of the Schwinger model factorizes
into a massive Fock space and a massless one. The condition (\ref{3.14})
only implies a restriction on the massless (conformally
invariant) sector which describes the ground state of the theory.
This restriction is equivalent to the corresponding BRST condition.
Indeed, the action associated with the Lagrangian (\ref{3.11})
is invariant under the BRST transformation
\bear\label{3.21}
&&2\sqrt\pi\delta\eta=-i \epsilon c_-\nonumber\\
&&\delta\chi_1=\epsilon c_-\chi_1,\ \delta\chi_2=0\nonumber\\
&&\delta c_-=\delta c_+=0\nonumber\\
&&\delta b_-=-\epsilon\chi_1^\dagger\chi_1-\frac{\epsilon}
{2\sqrt\pi}\partial_-
\eta,\
\delta b_+=0\ear
and
a similar transformation obtained by the substitutions  $\chi_1
\leftrightarrow \chi_2,\ c_\pm\leftrightarrow c_\mp$ and $b_\pm
\leftrightarrow b_\mp$.

These symmetries imply the conservation of the BRST currents,
\be\label{3.23}
\tilde J_\pm^{(B)}=c_\pm\tilde\Omega_\pm,\qquad \partial_\mp\tilde J
_\pm^{(B)}=0\ee
with $\tilde\Omega_\pm$ given by (\ref{3.13}). The condition
(\ref{3.14}) is seen to be equivalent to the BRST condition
\be\label{3.24}
\tilde Q_\pm|\Psi_0\rangle=0\ee
where $\tilde Q_\pm$ is the (nilpotent) charge associated with
the currents (\ref{3.23}), and $|\Psi_0\rangle$ labels the
ground states.

The matter and negative metric part of $\tilde\Omega_\pm$ separately
satisfy a Kac-Moody algebra with level $k=1$ and
$k=-1$, respectively.

As is well known, there exists an infinite set of solutions of
(\ref{3.24}).
This is quite unlike the case of $QCD_2$, where the vacuum
degeneracy is finite, as we now demonstrate.

\bigskip
{\it The $QCD_2$ vacuum}
\medskip

{}From our discussion in section 2 we conclude that the physical states of
$QCD_2$ are obtained by solving the BRST conditions
\be\label{4.1}
Q_\pm^{(B)}|\Psi\rangle=0,\qquad |\Psi\rangle\in{\cal H}_{phys}\ee
with identification of states differing by BRST exact states.
In (\ref{4.1}), $Q_\pm^{(B)}$ are the charges associated with
the BRST currents (\ref{2.8}) or, equivalently, (\ref{2.20}). We
shall preferably work with the currents (\ref{2.20}) of the
non-local formulation, where the BRST condition (\ref{4.1})
becomes a restriction on the conformally invariant sector of
${\cal H}_{phys}$ describing the ground states $|\Psi_0\rangle$
of $QCD_2$:
\be\label{4.2}
Q_\pm^{(B)}|\Psi_0\rangle=0\ee

The crucial observation now is that the solution of (\ref{4.2})
in the conformally invariant $f-gh-\tilde V$ sector of
(\ref{2.16}) is identical to the solution of the cohomology problem
of a level one $G/G$ coset WZW model, which corresponds to a
topological field theory. In the following we shall restrict ourselves
to $G=SU(2)$.

The constraints $\Omega_\pm$ involve the matter currents
\be\label{4.3}
J^a_\pm=\frac{1}{2}\bar\chi t^a\gamma_\pm\chi\ee
the negative metric-field currents
\be\label{4.4}
\tilde J^a_-=-\frac{1+C_V}{4\pi}tr\left(t^a\tilde V i\partial_-
\tilde V^{-1}\right),\ \tilde J^a_+=-\frac{1+C_V}{4\pi}tr\left(t^a
\tilde V^{-1}i\partial_+\tilde V\right)\ee
and the ghost currents
\be\label{4.5}
J_\pm^{a_{(gh)}}=f_{abc} b^b_\pm c^c_\pm\ee
Since the two light-cone components can be treated independently,
we shall omit the subscript $\pm$ from here on.

The solution of the BRST condition (\ref{4.2}) is constructed in
terms of the highest weight eigenstates $|J,\tilde J\rangle$ of the
charges associated with isospin-3 component of the currents (\ref{4.3})
and (\ref{4.4}) (zero modes in the corresponding Laurent expansion)
\be\label{4.6}
J^3_0|J,\tilde J\rangle=J|J,\tilde J\rangle,\ \tilde J_0^3
|J,\tilde J\rangle=\tilde J|J,\tilde J\rangle\ee
with the highest weight condition
\be\label{4.7}
J_0^{1+i2}|J,\tilde J\rangle=0,\ \tilde J_0^{1+i2}|J,\tilde J\rangle=0.\ee
We define our Fock space be requiring that the state $|J,\tilde J\rangle$
be annihilated by the ``positive'' frequency parts of the currents
(\ref{4.2})-(\ref{4.4}):
\be\label{4.8}
J^a_{n>0}|J,\tilde J\rangle=0,\quad \tilde J^a_{n>0}|J,\tilde J
\rangle=0\ee
\be\label{4.9}
c^a_{n>0}|J,\tilde J\rangle=0,\quad b^a_{n>0}|J,\tilde J
\rangle=0\ee
In (\ref{4.8}) and (\ref{4.9}) the subscript $n$ labels the modes in the
Laurent
expansion of the corresponding operators.
Since $b^a$ and $c^a$ are canonically conjugate fields, we further
require
\be\label{4.10}
b^a_0|J,\tilde J\rangle =0\ee
In order to obtain the states satisfying the BRST condition
(\ref{4.2}), in terms of the states $|J,\tilde J\rangle$ defined
above, we follow closely the work of ref. \cite{AGSYS} and make
use of the Wakimoto realization \cite{Wa} of the (level one)
Kac-Moody currents $J^a$ and $\tilde J^a$:
\bear\label{4.11}
&&J^+_n=a_n\nonumber\\
&&J_n^3=\sqrt{\frac{3}{2}}\phi_n+\sum_m a_md_{n-m}\nonumber\\
&&J_n^-=nd_n-\sqrt6\sum_{n,m} \phi_md_{n-m}\nonumber\\
&&-\sum_{m,n,k}d_md_k a_{n-m-k},\ear
where $a_n$ and $d_m$ are canonically conjugate pairs,
$[d_n,a_m]=\delta_{m+n}$
and
$[\phi_m,\phi_n]=m\delta_{m+n}$.
In the case of $U(1)$ (Schwinger model), $a_m=d_m=0$,
and $\sqrt{\frac{3}{2}}\phi_n\to\frac{1}{2\sqrt\pi}\partial\varphi$,
where the field $\varphi$ is defined in (\ref{3.15}).
An analogous decomposition in terms of $\tilde a_n, \tilde d_n$
and $\tilde\phi_n$ is made\footnote{Actually there are some technical
subleties,
for which we refer the reader to ref. \cite{AGSYS}.} for $\tilde J_n^\pm,
\tilde J_n^3$

Expressing the BRST charge in terms of the Wakimoto variables,
it can be decomposed into terms of given degrees by attributing
to the fields
$ c,d,\tilde d $ and $\phi^+
( h, a,\tilde a,\phi^-)$ the degree $+1(-1)$, where
$\phi^\pm$ are defined by
 $\phi^\pm=\frac{1}{\sqrt2}(\phi\pm i\tilde\phi)$. It turns
out that it suffices to study the states which are annihilated
by the operator $Q^{(0)}$ of lowest degree, which is nilpotent, as well
as quadratic in the fundamental excitations:
\be\label{4.15}
Q^{(0)}|\Psi_0\rangle=0\ee
\be\label{4.16}
Q^{(0)}=\sum_n c^-_{-n}a_n+2\sqrt3\sum_n c^3_{-n}\phi^-_n+
c_{-n}^+\tilde a_n\ee
The total current, as well as the zero'th mode of the energy
momentum tensor  (Virasoro operator $L_0$) correspondingly
take the form,
\be\label{4.17}
\Omega_0^3=J+\tilde J+\sum_n\left[:a_{-n}d_n:-:\tilde a_{-n}\tilde
d_n:-f_{3cb}:c^c_nb^b_{-n}:\right]\ee
\bear\label{4.18}
L_0&=&\frac{1}{3}[J(J+1)-\tilde J(\tilde J+1)]+\sum_{n\not=0}
\Bigl\{n:a_{-n}d_n\nonumber\\
&&+\tilde a_{-n}\tilde d_n+g_{bc}b^b_{-n}c^c_n]:+\phi^+_{-n}
\phi^-_n\Bigr\}\ear
As shown in ref. \cite{AGSYS}, the physical states must also
be annihilated by these operators:
\be\label{4.19}
\Omega^3_0|\Psi_0\rangle=0,\quad L_0|\Psi_0\rangle=0\ee
since $\Omega^3_0$ and $L_0$ turn out to be BRST exact. This
implies $\tilde J=-J-1$, as well as the absence of non-zero-
mode excitation, and allows one to write the physical
states as linear combinations of
\be\label{4.20}
|n_d,n_{\tilde a},n_+,n_-\rangle=(d_0)^{n_d}(\tilde a_0)^{n_{\tilde a}}
(c_0^+)^{n_+}(c_0^-)^{n_-}|J,-J-1\rangle\ee
Implementation of condition (\ref{4.15}) is then found to imply
\cite{AGSYS}
\be\label{4.21}
|\Psi_0\rangle=c_0^+|J,-(J+1)\rangle\ee
In order to completely classify the vacua, we still need to know
the values which $J$ can take. From the representation theory of
Kac-Moody algebras with central charge $k$ \cite{KK} one learns
that the allowed values of $J$ are finite and parametrized by
\bear\label{4.22}
2J+1&=&r-(s-1)(k+2)\nonumber\\
2J+1&=&-r+s(k+2)\ear
where $s=1,...,q$ and $r=1,...,p-1$, where $p$ and $q$ are
coprime and defined by $k+2=p/q$. For our case $k=1$. We
therefore conclude that $J=0,\frac{1}{2}$, that is, we have
a two-fold degeneracy of the ground state.

\medskip

Summarizing, we have shown, the BRST conditions implied restrictions on
the conformally invariant (vacuum) sector of $QCD_2$. By systematically
exploring these conditions, we have shown that for a given
chirality the vacuum of $SU(2)-QCD_2$ is two-fold degenerate.

The same conclusion is suggested by other, quite different considerations
based on the equivalence of the level $k,G/G$ model on the
space $\Sigma$ to   the so-called BF theories, which
in turn are equivalent to a Chern-Simons theory on $\Sigma\times S^1$
\cite{BT}. The result obtained in \cite{BT} for $G=SU(2)$ in
particular can be interpreted as the existence of $k+1$ states.
This agrees with the result (\ref{4.22}) obtained from representation
theory, which for $k$=integer$(s=1)$ reduces to just one condition,
$2J+1=r, r=1,...,k+1$. In our case, $k=1$.

In the case $G=SU(N)$ we also
expect a discrete, growing number of vacua, in accordance with the
results of ref. \cite{BT}. In the case of the Schwinger model \cite{Schw}
and its generalization to the Cartan subalgebra of $U(N)$ \cite{BRSS},
the vacuum is infinitely degenerate. In the $U(1)$ case this is
known since the work of Lowenstein and Swieca \cite{LS}; in the
framework of section 4, these vacua are given by $|\Psi_0\rangle
=|J,-J\rangle$, and the infinite degeneracy is a consequence of
$J$ taking all integer values.

In the non-local formulation of $QCD_2$ we have seen that the massive
sector (described by $Z_\beta$) completely separates from the
conformally invariant one (describing the vacuum sector). In
\cite{AA4} the $S$-matrix has been computed up to pole factors
describing the bound states. These factors may differ according
to the choice of vacuum.

\bigskip
{\it Acknowledgment}
\medskip

One of the authors (E.A.)
is grateful to O. Aharony, L. Alvarez Gaum\'e, and G. Thompson
for some useful discussions and correspondence. He would als
like to thank the Alexander von Humboldt Foundation for financial
support making this collaboration possible.

\end{document}